\documentclass{article}
\usepackage{arxiv}
\pdfoutput=1
\usepackage[utf8]{inputenc} 
\usepackage[T1]{fontenc}    
\usepackage{hyperref}       
\usepackage{url}            
\usepackage{booktabs}       
\usepackage{amsfonts}       
\usepackage{nicefrac}       
\usepackage{microtype}      
\usepackage{subfiles}
\usepackage{amsmath}
\usepackage{enumitem}
\usepackage{multirow}
\usepackage{xcolor}
\usepackage{soul}
\usepackage{subfiles}
\usepackage{graphicx}
\usepackage[superscript,biblabel]{cite}
\usepackage{amssymb}

\usepackage{lineno}
\usepackage[T1]{fontenc}
\usepackage{amsmath}
\usepackage{enumitem}
\usepackage{multirow}
\usepackage{xcolor}
\usepackage{soul}
\usepackage{subfiles}
\usepackage{authblk}
\usepackage{dirtytalk}
\usepackage{subcaption}
\title{Exploiting generative self-supervised learning for the assessment of biological images with lack of annotations: a COVID-19 case-study}

\begin{document}

\author[1,$\mathsection$,$\ddagger$]{Alessio~Mascolini}
\author[2,3,$\mathsection$]{Dario~Cardamone}
\author[1,*]{Francesco~Ponzio}
\author[1,$\dagger$]{Santa~{Di~Cataldo}}
\author[4,$\dagger$]{Elisa~Ficarra}
\affil[1]{Politecnico di Torino, Dept. of Control and Computer Engineering, Torino, Italy}
\affil[2]{Università degli Studi di Torino, Dept. of Physics, Torino, Italy}
\affil[3]{Monoclonal Antibody Discovery (MAD) Lab, Fondazione Toscana Life Sciences, Siena, Italy}
\affil[4]{Università di Modena and Reggio Emilia, Dept. of Engineering "Enzo Ferrari", Modena, Italy}
\affil[$\mathsection$]{These authors contributed equally to this work}
\affil[$\dagger$]{These authors contributed equally to this work}
\affil[$\ddagger$]{Research supported with Cloud TPUs from Google's TensorFlow Research Cloud (TFRC)}
\affil[*]{Corresponding author}

\maketitle

\begin{abstract}
Computer-aided analysis of biological images typically requires extensive training on large-scale annotated datasets, which is not viable in many situations.
In this paper we present GAN-DL, a Discriminator Learner based on the StyleGAN2 architecture, which we employ for self-supervised image representation learning in the case of fluorescent biological images.
We show that Wasserstein Generative Adversarial Networks combined with linear Support Vector Machines enable high-throughput compound screening based on raw images. We demonstrate this by classifying active and inactive compounds tested for the inhibition of SARS-CoV-2 infection in VERO and HRCE cell lines. In contrast to previous methods, our deep learning based approach does not require any annotation besides the one that is normally collected during the sample preparation process.
We test our technique on the RxRx19a Sars-CoV-2 image collection. The dataset consists of fluorescent images that were generated to assess the ability of regulatory-approved or in late-stage clinical trials compound to modulate the in vitro infection from SARS-CoV-2 in both VERO and HRCE cell lines.
We show that our technique can be exploited not only for classification tasks, but also to effectively derive a dose response curve for the tested treatments, in a self-supervised manner. Lastly, we demonstrate its generalization capabilities by successfully addressing a zero-shot learning task, consisting in the categorization of four different cell types of the RxRx1 fluorescent images collection.
\end{abstract}


\section*{Introduction}
\label{sec:intro}
A good feature representation is a key aspect for any visual recognition task. Thanks to their inherent capability to discover hidden data structures, as well as to extract powerful features representation, Convolutional Neural Network (CNNs) have become the fundamental building blocks in most computer vision applications. Nevertheless, much of their recent success lies in the existence of large labeled datasets: CNNs are data-hungry supervised algorithms, and thus supposed to be fed with a large amount of high quality annotated training samples~\cite{liu2020SSRL_review}. 

However, associating labels to a massive number of images to effectively train a CNN may be extremely problematic in a number of real-world applications. Significant examples are the medical and computational biology domains, where  
image annotation is an especially cumbersome and time-consuming task that requires solid domain expertise and, more often than not, necessitates consensus strategies to aggregate annotations from several experts to solve class variability problems~\cite{retinal, wallace2020extending_SSRL,ponzio2021w2wnet}. Moreover, biological systems are affected by multiple sources of variability that make the definition of a supervised task impractical, as they require to discover new effects that were not observed during the generation of the training set. On the other hand, a considerable amount of literature focused on machine learning systems, especially CNNs, able to adapt to new conditions without needing a large amount of high-cost data annotations. 
This effort includes advances on transfer learning, domain adaptation, semi-supervised learning and self-supervised representation learning \cite{liu2020SSRL_review}. 

Among the context shifting techniques, the self-supervised representation learning (SSRL) paradigm has recently received an increasing attention in the research community. Yann LeCun, invited speaker at AAAI 2020 conference \cite{AAAI}, has defined the SSRL as \say{the ability of a machine to predict any parts of its input from any observed part}. In other words, SSRL can be realized by contextualizing a supervised learning task in a peculiar form (known as \emph{pretext task}) to predict only a subset of the information using the rest to drive the decision process. Although the pretext task guides the learning through a supervised loss function, the performance of the model on the pretext is irrelevant, as the actual objective of SSRL is to learn an intermediate representation capable of solving a variety of practical downstream tasks. Popular SSRL pretext tasks are rotation, jigsaw, instance discrimination and autoencoder-based methods (colorization, denoising, inpainting)~\cite{liu2020SSRL_review, wallace2020extending_SSRL}. 

Nevertheless, current literature has primarily exploited SSRL on general category object classification tasks (e.g. ImageNet classification)~\cite{liu2020SSRL_review, wallace2020extending_SSRL}. 
Surprisingly, there has been very little attention on how to extend SSRL methodologies to other domains like computational biology or medicine, which paradoxically are among the ones that are most affected by the lack of labeled training data~\cite{wallace2020extending_SSRL}. 
In this sense, for contexts distant from the standard natural image benchmarks, finding a pretext task capable of learning a reliable and robust data representation is of particular concern. A recent longitudinal investigation by Wallace et al.~\cite{wallace2020extending_SSRL} shows how traditional SSRL feature embedding fails in several biological downstream tasks. The authors suggest that the absence of canonical orientation, coupled with the textural nature of the problems, prevents SSRL popular methods from learning a pertinent representation space. They conclude that finding an optimal SSRL feature embedding for fine-grained, textural and biological domains is still an open question. 

Motivated by the findings of Wallace et colleagues ~\cite{wallace2020extending_SSRL}, in this study we propose \emph{GAN Discriminator Learner} (GAN-DL), a SSRL framework based on the discriminator of a state-of-the-art Generative Adversarial Network (GAN), namely the StyleGAN2 model~\cite{styleGAN2}. The training of the StyleGAN2 backbone is based on the competition of a generator and of a discriminator, that does not require any task-specific annotation. We specifically seek a SSRL-based featurization methodology capable of learning a reusable and application-independent image representation that is exploitable in complex biological domains which embody the textural as well the fine-grained patterns that typically lead the traditional SSRL techniques to failure \cite{liu2020SSRL_review}. 

To characterize our framework, we focus on a particularly complex biological case-study, that is COVID-19 drug discovery, exploiting two recently released fluorescence microscopy datasets: (i) the RxRx19a, a morphological imaging dataset that is specific of COVID-19~\cite{RxRx19}; (ii) the RxRx1, a non-COVID related collection of fluorescent microscopy images~\cite{RxRx1} (a more detailed description will follow).

\begin{figure}[h!]
\centering
\includegraphics[width=0.8\textwidth]{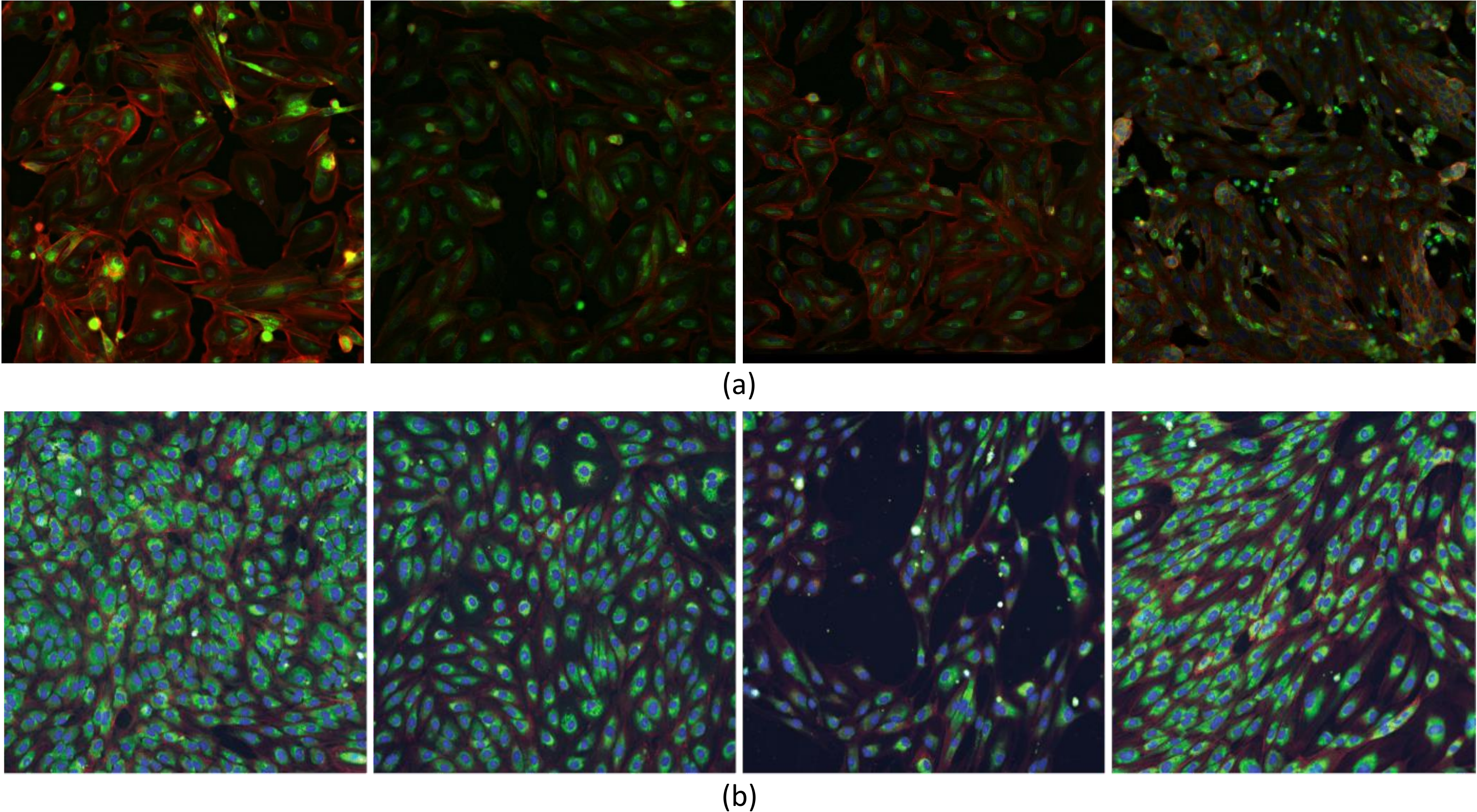}
\caption{Representative examples of RxRx19a (a) and RxRx1 datasets (b).}
\label{fig:images_examples}
\end{figure}

\figureautorefname~\ref{fig:images_examples} gathers some representative images taken from RxRx19a (a) and RxRx1 (b) datasets: the image content largely diverges from those of typical SSRL benchmarks (e.g. the ImageNet). Thus, such datasets perfectly embody those features (absence of canonical orientation, fine-grained content, textural nature) that make difficult, or even not solvable, the classical SSRL pretext tasks as described in the work by Wallace and colleagues \cite{wallace2020extending_SSRL}.

Besides the imaging data, for the RxRx19a benchmark, a transfer learning-based image embedding is also online accessible \cite{RxRx19, RxRx2}. Such embedding is taken as baseline comparison to prove the goodness of our approach, and referred to as \emph{baseline} in the rest of the manuscript. 

\begin{figure}[h!]
\includegraphics[width=.9\textwidth]{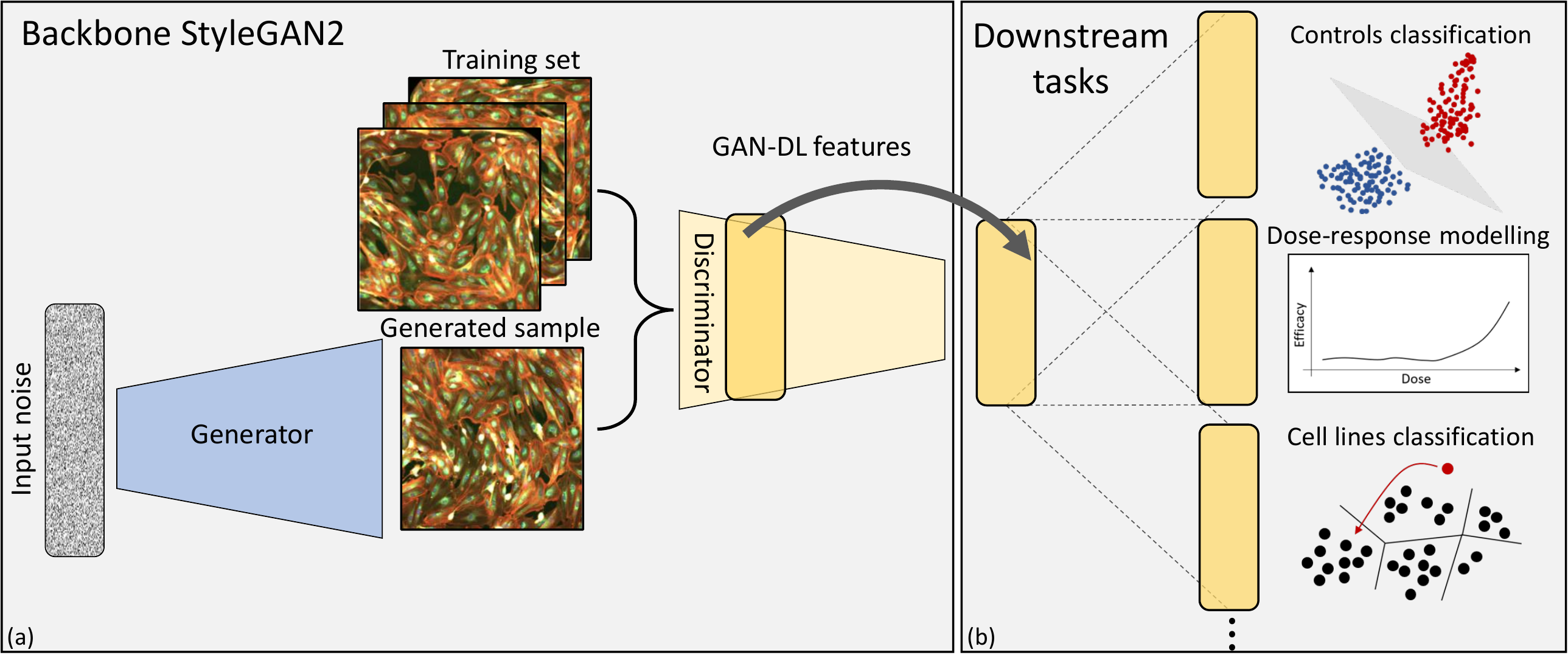}
\centering
\caption{Overview of GAN-DL self-supervised representation learning framework, whose pretext task consists in the adversarial game between the generator and the discriminator of the backbone StyleGAN2 (a). The discriminator's features are exploited to several downstream tasks (b): (i)~Controls classification - classification of active and inactive compounds against SARS-CoV2 in two different cell models; (ii)~Dose-response modelling - disease-associated profiling from raw microscopy images; (iii)~Cell lines classification - zero-shot learning classification task consisting in categorizing four different cell types.}
\label{fig:system_overview}
\end{figure}

Our main contributions are the followings:
\begin{enumerate}[label=\roman*.]
    \item We propose GAN-DL, a fully SSRL-based approach to characterize complex biological case studies. Up to our knowledge, our work is the first to employ SSRL in challenging, real-world biological applications.
    
    \item GAN-DL leverages the features of the discriminator of a StyleGAN2 model~\cite{styleGAN2} to learn the RxRx19a data distribution without needing any specific image labelling (see \figureautorefname~\ref{fig:system_overview}(a)). Our procedure has its own foundation in a pretext task which does not require any modification of the original data: the adversarial game between the GAN's generator $G$ and discriminator $D$ ($G$ tries to fool $D$, which in turn tries to identify real data from the fake ones created by $G$). We show that GAN-DL, leveraging the pretext of creating diverse and realistic images, is fully independent from those morphological and geometrical aspects which hampered the adoption of canonical SSRL techniques in medical and biological domains~\cite{wallace2020extending_SSRL}. 
    
    For the sake of clarity, it must be specified that we are not the first to exploit feature extraction based on GAN's discriminator. Such procedure was firstly introduced by Radford et al. in 2017~\cite{DCGAN}. After this pioneering study, discriminator-based feature extractors have been also exploited in other specific applications~\cite{lin2017_discriminator_fe1, zhang2018_discriminator_fe2}. More recently, Mao et al \cite{Mao} showed that the effectiveness and robustness of discriminator features strongly depends on avoiding mode collapse in the network. This motivated our choice of using StyleGAN2 \cite{styleGAN2} as backbone: the Wasserstein family of GANs, among which StyleGAN2, are known to be particularly resistant to this phenomenon \cite{arjovsky2017wasserstein, zhang2018_discriminator_fe2}. 
    
    Up to our knowledge, we are instead the first to exploit GAN's discriminator features in an extremely challenging biological context, coupled with high resolution microscopy images. In such context, we propose and fully investigate an embedding capable not only of proficiently managing downstream classification tasks, but also of separating multiple unrelated features at once along different axis of the latent space.
    
    \item GAN-DL significantly deviates from the baseline featurization method proposed by Cuccarese et al.~\cite{baseline} and released together with the RxRx19a benchmark. As a matter of fact, the authors proposed a classic transfer-learning approach featuring a deep network trained from scratch on the RxRx1~\cite{RxRx1}, a very large dataset that is similar in terms of imaging technology and content to their final application, the RxRx19a \cite{RxRx19}, but with a much higher annotated information content. The necessity of a pre-training phase leveraging about 300GB of annotated microscopy images puts serious limitations to the applicability of such method in other contexts affected by scarcity of labelled data. Conversely, as above-mentioned, GAN-DL does not need any labelling.
   
    \item To assess GAN-DL's ability to solve different downstream tasks, we evaluate our method on the classification of active and inactive compounds against SARS-CoV2 in two different cell lines (see \figureautorefname~\ref{fig:system_overview}(b)). We show that GAN-DL: (i)~outperforms the classical transfer learning approach consisting of a CNN pre-trained on ImageNet; (ii)~is comparable to the baseline method in terms of accuracy, even though it was not purposely trained for the downstream tasks; (iii)~is able to model disease-associated profiles from raw microscopy images, without the use of any purposely labelled data during the training.
    \item Finally, to assess the generalization capability of our method, we exploit the GAN-DL embedding learnt on RxRx19a in a zero-shot learning task consisting in categorizing the four different cell types of the RxRx1 benchmark\cite{RxRx1}: human liver cancer cells (HEPG2), human umbilical vein endothelial cells (HUVEC), retinal pigment epithelium cells (RPE) and human bone osteosarcoma epithelial cells (U2OS).
    \end{enumerate}

The remaining part of the paper proceeds as follows: Results \sectionautorefname~reports our experimental results; Discussion \sectionautorefname~outlines and discusses our findings; finally Methods \sectionautorefname~provides the description of our GAN-DL's implementation and the details about the proposed biological applications.

\section*{Experimental Results}
\label{sec:results}
\label{sec:latent_space}
Our experiments specifically seek an answer to two main research questions: (i) is it possible to learn an accurate and reliable image featurization, able to encode and describe biological relevant information, leveraging a self-supervised pretext task?; (ii) up to which extent the learned biological information can be reused in a different context? To answer such questions, we have put into effect the properties of GAN-DL featurization in the following experiments.

\subsection*{Visualizing GAN-DL's representation learning capability}
\label{results_clusters}

To characterize the representation capability of the proposed SSRL featurization methodology, we evaluate GAN-DL on the RxRx19a dataset, which gathers experimental data in the form of cellular imagery to investigate potential therapeutic treatments for COVID-19. Specifically, RxRx19a evaluates a library of 1,670 approved and referenced compounds in an unbiased, image-based screening study involving two different cell lines: the primary human renal cortical epithelial cells (HRCE) and the African green monkey kidney epithelial cells (VERO). Both the cell lines have been infected in-vitro with wild-type SARS-CoV2, and incubated 96 hours before fixation, staining and image acquisition\cite{rxrx}. 

Two suitable control groups have been designed to assess compounds specificity. The first one, referred to as positive control group ($C^+$) in the rest of the manuscript, consists of uninfected \emph{mock-treated} cells, namely samples treated with culture medium or a solvent without any active compound nor any infection. The second control group, hereafter referred to as negative control group ($C^-$), is made up of cells infected in vitro by wild-type SARS-CoV-2 virus and not treated with any compounds.


The remaining part of RxRx19a consists in the actual drugs screening, where the infected cells are treated with compounds at different concentration.
It is reasonable to assume that effective compounds will be able to inhibit the infection and maintain a cell viability comparable to the positive controls.

In the RxRx19a compound screening setting, only the positive and negative control images can be univocally associated with either \emph{live} or \emph{dead} labels. The remaining part of the samples, which is the vast majority of the dataset, is, in this regards, unlabelled. The large amount of unlabelled data, coupled with the textural and fine-grained aspect of the images, makes RxRx19a a very challenging case-study and a perfect candidate to assess our proposed SSRL methodology.

As \figureautorefname~\ref{fig:system_overview} suggests, GAN-DL embedding consists of a high-dimensional feature vector (512 features, see Materials and Methods for details).  Hence, to assess and interpret its inherent capability of learning a genuine featurization, we need to define a projection space able to allow some degrees of visualization of the data structure. Hence, we promote the explainability of the projection procedure defining: 
\begin{enumerate}
    \item the \emph{effectiveness-space} $\textbf{E}^2$, a two-dimensional space that represents the treatment effectiveness of the tested compounds on the \emph{On-perturbation} and \emph{Off-perturbation} axes. The \emph{On-perturbation} axis of $\textbf{E}^2$ must catch the variability between $C^+$ and $C^-$ deriving from the expected different cell viability due to the viral infection. Ultimately, the $C^+$ and $C^-$ control groups should be well-separated on such direction. Accordingly, the remaining samples of the RxRx19a dataset should cluster towards $C^+$ and $C^-$ depending on the effectiveness of the given compound: samples characterized by alive cells, namely effective compounds, should be grouped around $C^+$; samples characterized by dead cells, i.e. ineffective compounds, around $C^-$. The scalar projection of the features of a given sample on the \emph{On-perturbation} axis defines the \emph{efficacy score} which testifies whether or not the given compound is effective against in-vitro SARS-CoV-2. More details on how to construct \textbf{$E^2$} and compute the \emph{efficacy score} will follow later in this section. By contrast, the \emph{Off-perturbation} axis of $\textbf{E}^2$ gathers the remaining variability of the data, not attributable to the effectiveness of the compounds.
    \item the \emph{cell lines-space} $\textbf{C}^2$, a two-dimensional space whose \emph{On-perturbation} axis captures morphological and functional data properties capable of grouping the samples into two clusters stemming from the two different cell lines used in the screening: HCRE and VERO cells. Similarly to the previous case, the \emph{Off-perturbation} axis of $\textbf{C}^2$ embodies those variability not ascribable to the two different cellular models considered. More details on how to construct $\textbf{C}^2$ will follow later in this section.
\end{enumerate}

    
Concerning $\textbf{E}^2$, a similar concept of \emph{On/Off-perturbation} axes was firstly reported in the work by Cuccarese at al.\cite{baseline}, respectively corresponding to the direction connecting the barycenters of the clusters of control conditions (\emph{On-perturbation}), and its orthogonal (\emph{Off-perturbation}) in the embedding space. This can be intuitively described as a projection that explicitly divides the variation of the data along a direction of interest (\emph{On-perturbation}) from all the remaining variations, that are grouped in the orthogonal hyperplane (\emph{Off-perturbation}). Here we expand and automatize this definition by exploiting a data-driven solution. More specifically, we leverage a linear Support Vector Machine (SVM) trained to classify: (i) $C^+$ versus $C^-$; (ii) HRCE versus VERO cells. In both the cases, the \emph{Off-perturbation} axis is defined as the separation hyperplane fitted by the SVM, while the \emph{On-perturbation} one is its normal. Thus, leveraging the \emph{On/Off} perturbation directions, we can define the aforementioned two-dimensional reference spaces $\textbf{E}^2$ and $\textbf{C}^2$, related to first and second classification task, respectively. The scalar projection of the features embedding on such spaces produces two components, exploited on one hand to visually interpret the data structure through point cloud scatter plots, on the other hand to derive dose-response curves for the tested compounds, as shown later in this section. Lastly, for a better readability of the plots, we have scaled the \emph{On-perturbation} axis of both $\textbf{C}^2$ and $\textbf{E}^2$ in a $[-1, 1]$ range (min-max feature scaling). Accordingly, the \emph{Off-perturbation} axis has been zero-centred.

\begin{figure}[ht]
     \centering
     \begin{subfigure}[b]{0.49\textwidth}
         \centering
         \includegraphics[width=\textwidth]{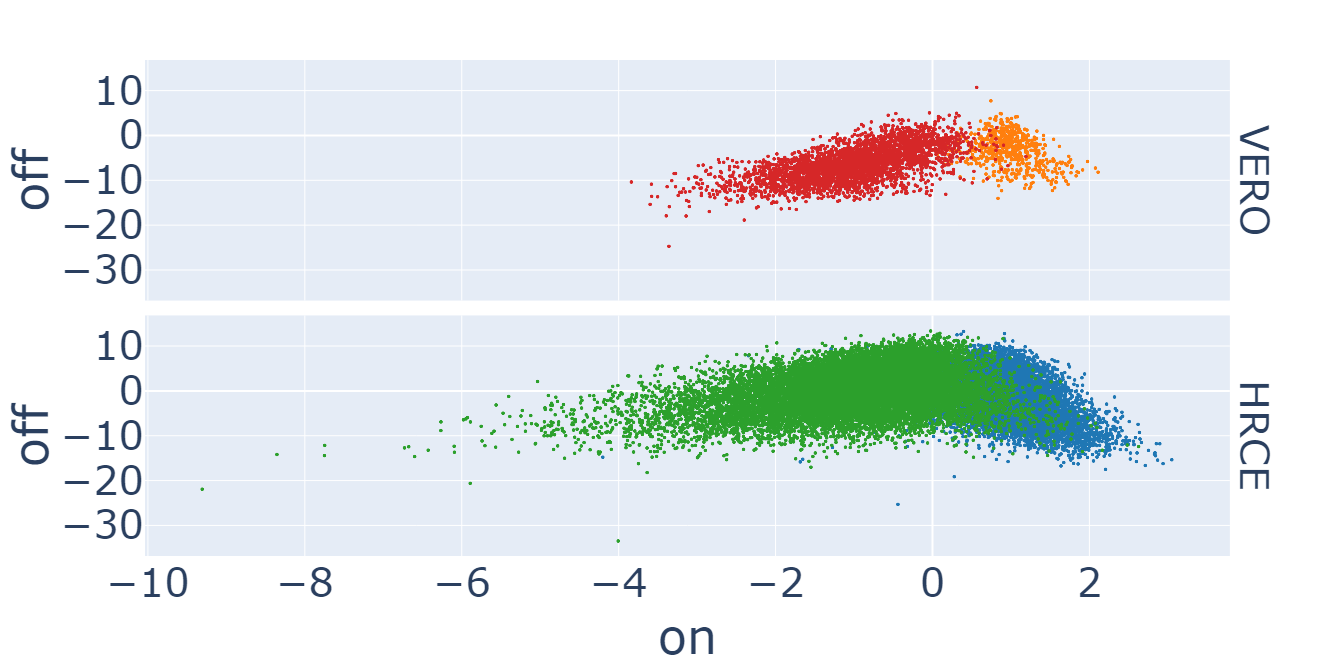}
         \caption{GAN-DL's embedding, $\textbf{E}^2$ space}
     \end{subfigure}
     \hfill
     \begin{subfigure}[b]{0.49\textwidth}
         \centering
         \includegraphics[width=\textwidth]{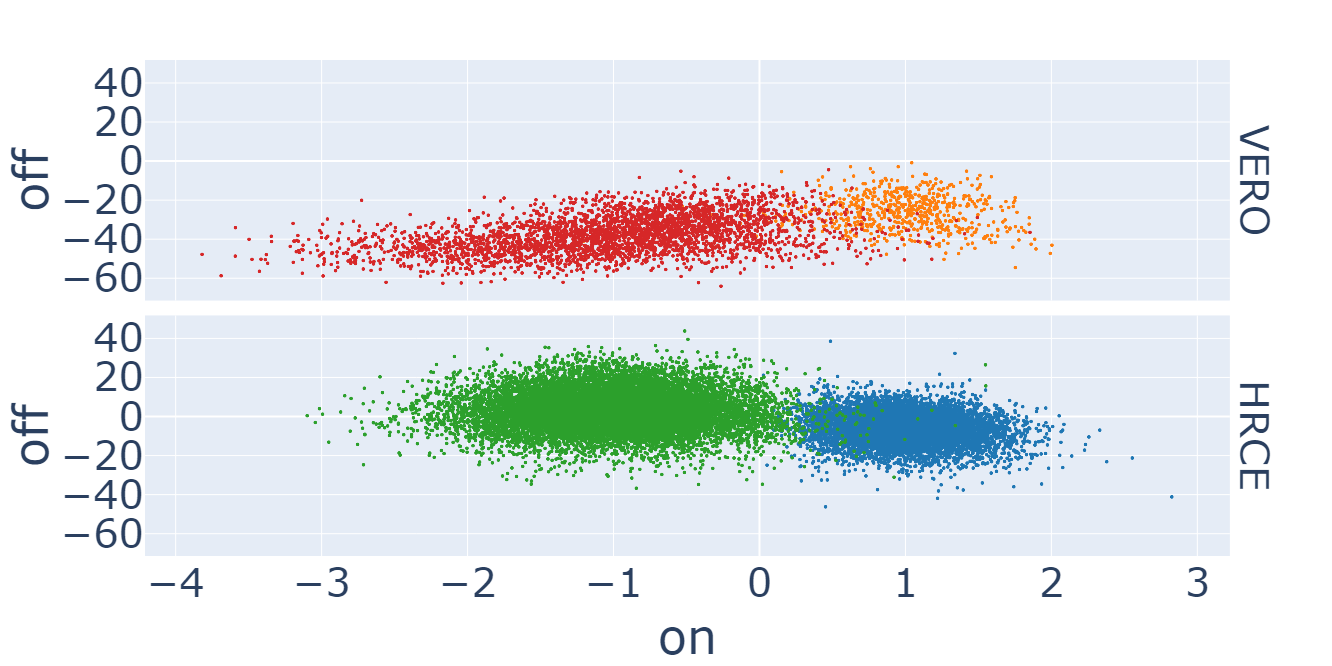}
         \caption{Baseline embedding, $\textbf{E}^2$space}
     \end{subfigure}
     
     \begin{subfigure}[b]{0.49\textwidth}
         \centering
         \includegraphics[width=\textwidth]{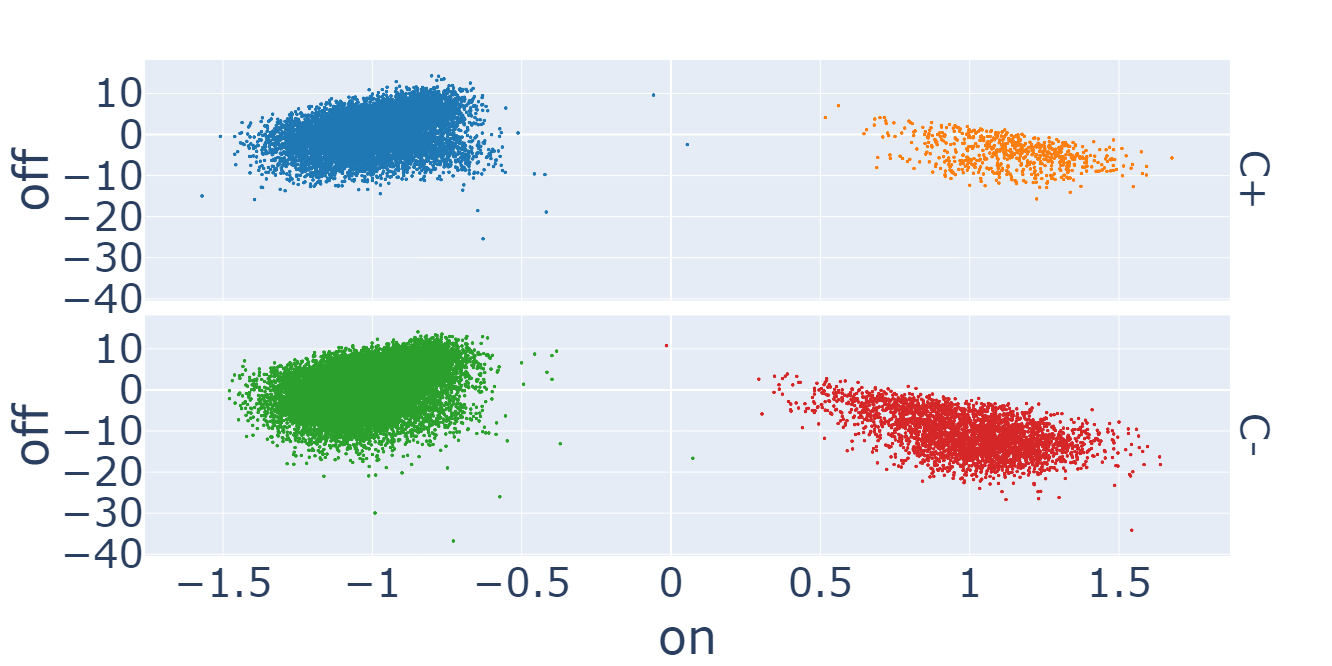}
         \caption{GAN-DL's embedding, $\textbf{C}^2$ space
         \label{fig:clustering_b}}
     \end{subfigure}
     \hfill
     \begin{subfigure}[b]{0.49\textwidth}
         \centering
         \includegraphics[width=\textwidth]{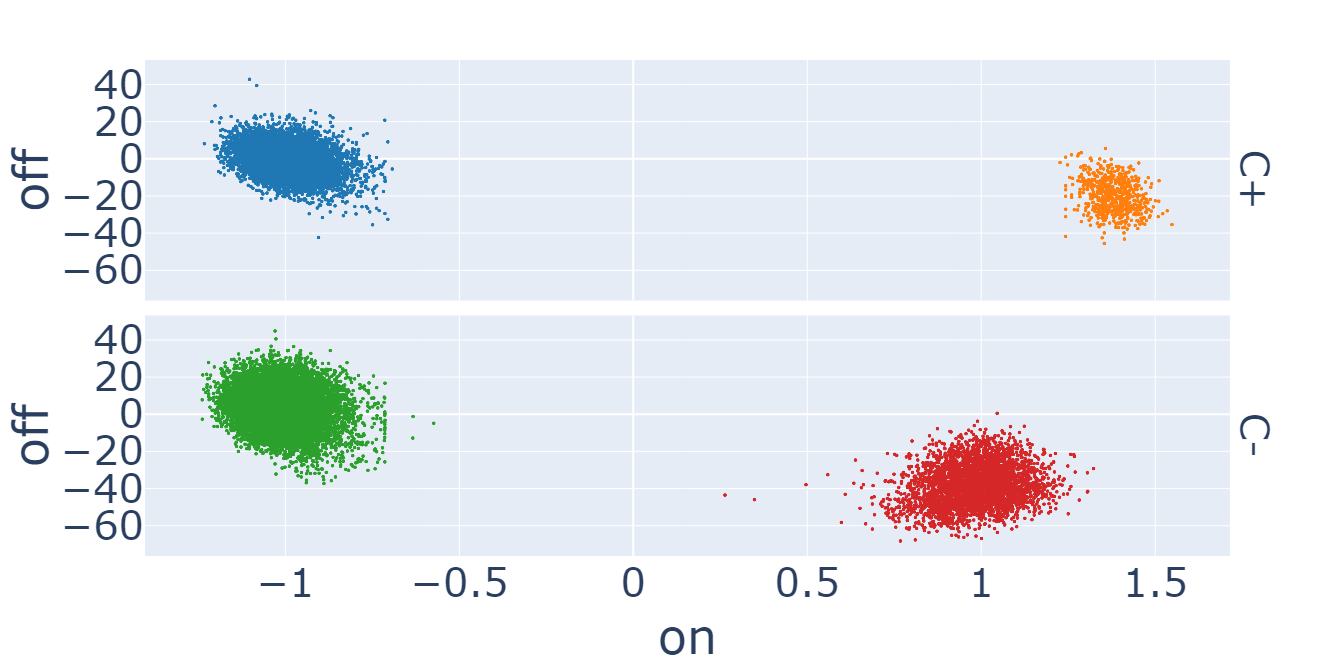}
         \caption{Baseline embedding, $\textbf{C}^2$ space}
     \end{subfigure}
     \vspace{10pt}
     \begin{subfigure}[b]{0.49\textwidth}
         \centering
         \includegraphics[width=\textwidth]{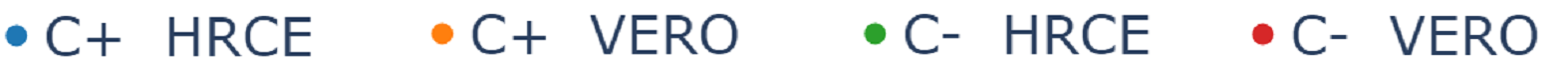}
     \end{subfigure}
    \caption{Scatter plots of GAN-DL's (left column) and baseline (right column) embeddings of the RxRx19a dataset projected onto the $\textbf{E}^2$(a-b) and $\textbf{C}^2$(c-d) axes.}
    \label{fig:clustering}
\end{figure}

The plots gathered in the first row of \figureautorefname~\ref{fig:clustering} compare our GAN-DL's embedding (a) with the baseline embedding~\cite{baseline}(b) in the $\textbf{E}^2$ projection space, where we expect a degree of separation between $C^-$ and $C^+$. The analysis is performed considering the positive and negative controls grouped by cell type. Hence, different colors identify $C^-$ and $C^+$ for the two distinct cell lines: blue and orange for the positive controls of HRCE and VERO cell lines, respectively, green and red for the corresponding negative controls. As it can be gathered from the degree of separation between $C^-$ and $C^+$ on the $\textbf{E}^2$ projection space, both the embeddings behave coherently in separating mock-treated samples from those where the virus was active. A quantitative comparison in terms of degree of separation between $C^-$ and $C^+$ is presented in the following subsection. 

The second row of \figureautorefname~\ref{fig:clustering} shows GAN-DL featurization (c) and the baseline featurization (d) projected onto the $\textbf{C}^2$ space, where we expect a certain degree of separation between distinct cell types, irrespective of whether $C^-$ or $C^+$ are considered. Same as in the previous experiment, results are reported separately for the two cell lines. Here HRCE are represented with blue ($C^+$) and green ($C^-$) colors, while VERO with orange ($C^+$) and red ($C^-$), respectively. Even in this case, the plots demonstrate that GAN-DL is able to caught the inherent variability of the two cell lines, in a comparable way to the transfer-learning baseline. 

\subsection*{Assessing the linear separability of the controls}

Leveraging the information content of our featurization, we quantitatively assess the accuracy on two downstream tasks: (i)~the categorization of $C^-$ versus $C^+$ and (ii)~the classification of HRCE and VERO cells.

For both the classification tasks, we compare a soft margin linear SVM built on top of our GAN-DL embedding with two other equivalent SVMs, respectively built (i)~on top of the baseline featurization, and (ii)~on top of the embedding of a DenseNet CNN model, pre-trained on ImageNet and fine-tuned respectively on the controls classification, and on the cell lines categorization of RxRx19a dataset. Note that for both the classification tasks, only images belonging to the control have been exploited to fine-tune the ImageNet-based embeddings, as they are the only samples associated to labels. We specifically select a DenseNet for the sake of a fair comparison, as it is also the backbone architecture of the baseline methodology~\cite{baseline}. 

The first two lines of \tableautorefname~\ref{tab:dowstream_task} report the classification accuracy values of the two classification tasks (for the first one, $C^-$ versus $C^+$, the two cellular lines are merged into the same dataset). 
From the reported values we can observe that GAN-DL provides informative features for both $C^-$ versus $C^+$ categorization (91.4\% accuracy) and cell lines recognition (100\% accuracy). The baseline, that leverages the RxRx1\cite{RxRx1} dataset as transfer learning source domain, outperforms GAN-DL of just 5\% in term of $C^-$ versus $C^+$ classification accuracy, and is equivalently 100\% accurate in the other task. This is a remarkable result for GAN-DL, given that no pre-training on a similar annotated context was performed. Lastly, GAN-DL outperforms by a large margin (respectively, by 26\% and 14\% for the two tasks) the traditional transfer learning solution based on ImageNet pre-training and following dataset-specific fine-tuning. 

The last two lines of \tableautorefname~\ref{tab:dowstream_task} report again the accuracy of the $C^-$ versus $C^+$ categorization task, this time separated by the cellular models HRCE and VERO. For all the considered embeddings, we can observe that the accuracy is higher when the cell lines are separated. Nonetheless, this variation is quite contained for our solution, suggesting that the learnt representation is reasonably general irrespective of the cellular model. More specifically, GAN-DL shows an accuracy of 92.44\% and 99.93\% for respectively HRCE and VERO, against the 91.4\% obtained with the two lines considered together. The baseline, on the other hand, shows an accuracy of 99.28\% and 100\% for respectively HRCE and VERO, against the 95.81\% for the two merged cell lines. We can again observe that the ImageNet pre-trained solution reported a much higher accuracy difference: 84.09\% and 84.53\% against 65.31\%.


\begin{table}[]
\centering
\footnotesize
\caption{Classification accuracy on the downstream tasks.}
\label{tab:dowstream_task}
\begin{tabular}{cccc}
\hline
\multicolumn{1}{l}{}    & \textbf{GAN-DL} & \textbf{baseline {[}14{]}} & \multicolumn{1}{l}{\textbf{ImageNet pre-trained CNN}} \\ \hline
\textit{$C^+$ vs $C^-$} & 91.4 \%         & 95.81 \%                    & 65.31\%                                                 \\
\textit{HRCE vs VERO}   & 100.0 \%         & 100.0 \%                    & 85.52\%                                                    \\ 
\hline
\textit{C+ vs C- (HRCE only)} & 92.44 \% & 99.28 \% & 84.09 \% \\
\textit{C+ vs C- (VERO only)} & 99.93 \% & 100 \% & 84.53 \% \\
\hline
\end{tabular}
\end{table}

\subsection*{Automatically deriving dose-response curves from image data}
As discussed in the previous subsection, GAN-DL can successfully address controls categorization and cellular typing tasks. In this section, we show how GAN-DL's representation can explicitly describe salient and biologically relevant data attributes, related to the efficacy of the different compounds tested in the RxRx19a screening initiative. For this purpose, we automatically derive the dose-responce of all the 1,672 screened compouds solely from raw image data and exploiting the GAN-DL's featurization.

\begin{figure}[p!]
\centering

    \vspace*{-1cm}
    \begin{subfigure}[b]{0.49\textwidth}
         \includegraphics[width=\textwidth]{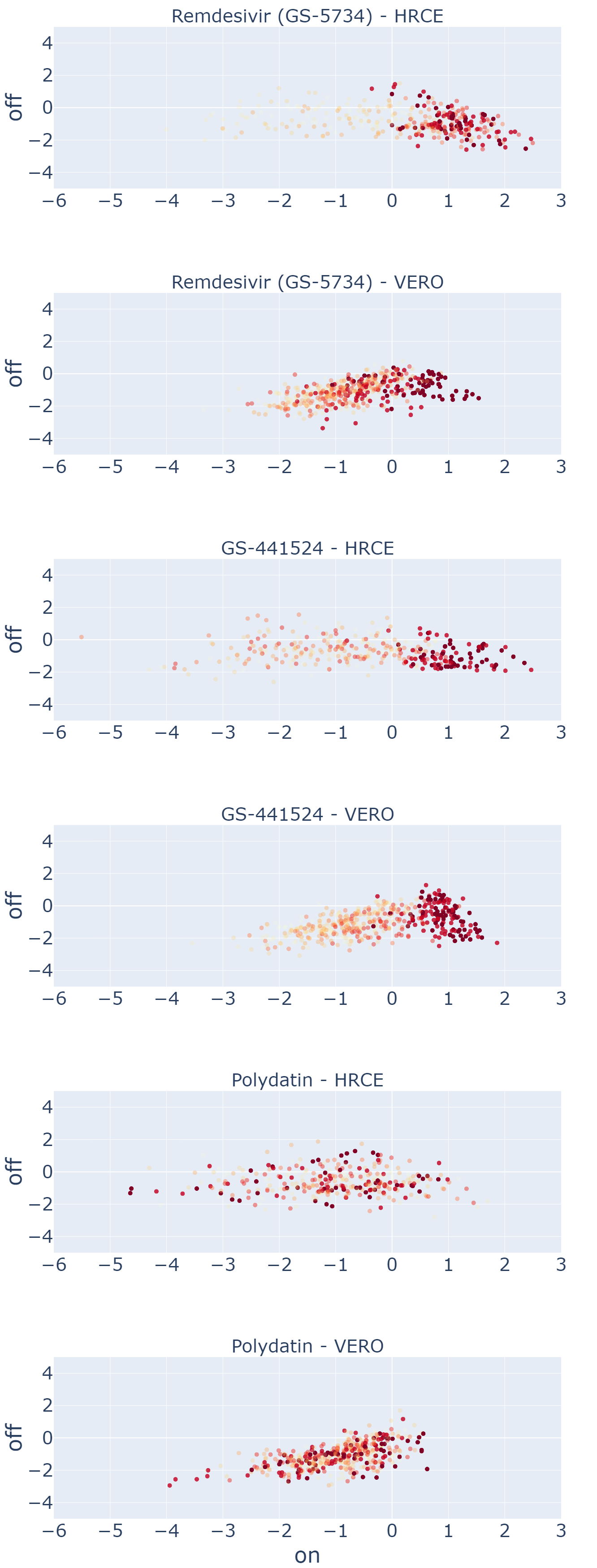}
         \caption{GAN-DL embeddings}
     \end{subfigure}
     \hfill
     \begin{subfigure}[b]{0.49\textwidth}
         \includegraphics[width=\textwidth]{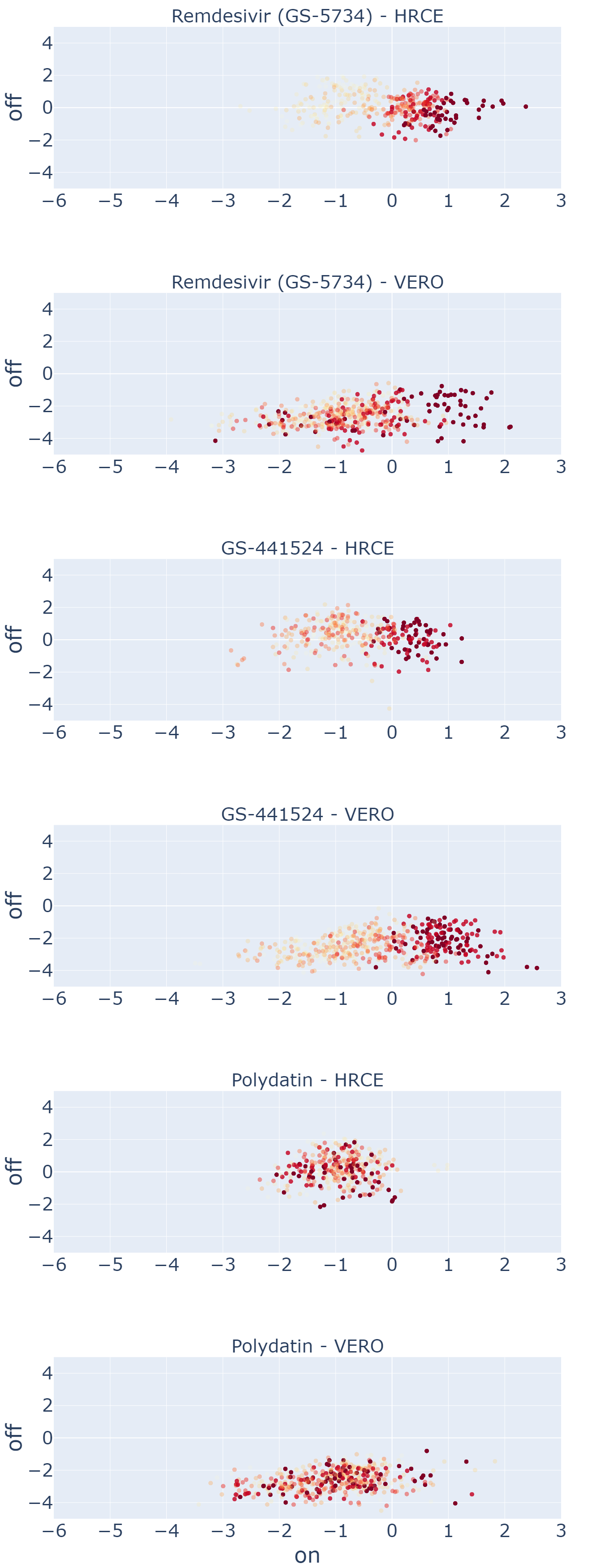}
         \caption{Baseline embeddings}
     \end{subfigure}
     
    \begin{subfigure}[b]{0.3\textwidth}
         \includegraphics[width=\textwidth]{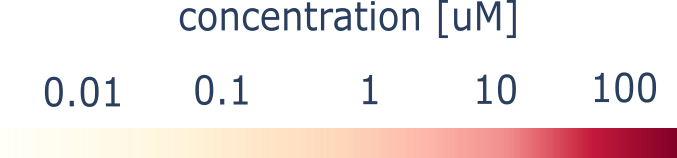}
     \end{subfigure}
 \caption{Drug effectiveness as a function of concentration, obtained using our GAN-DL (a) and the baseline embedding (b).}
\label{fig:dose_response_scatter}
\end{figure}

\begin{figure}[h]
\centering
\includegraphics[width=0.85\textwidth]{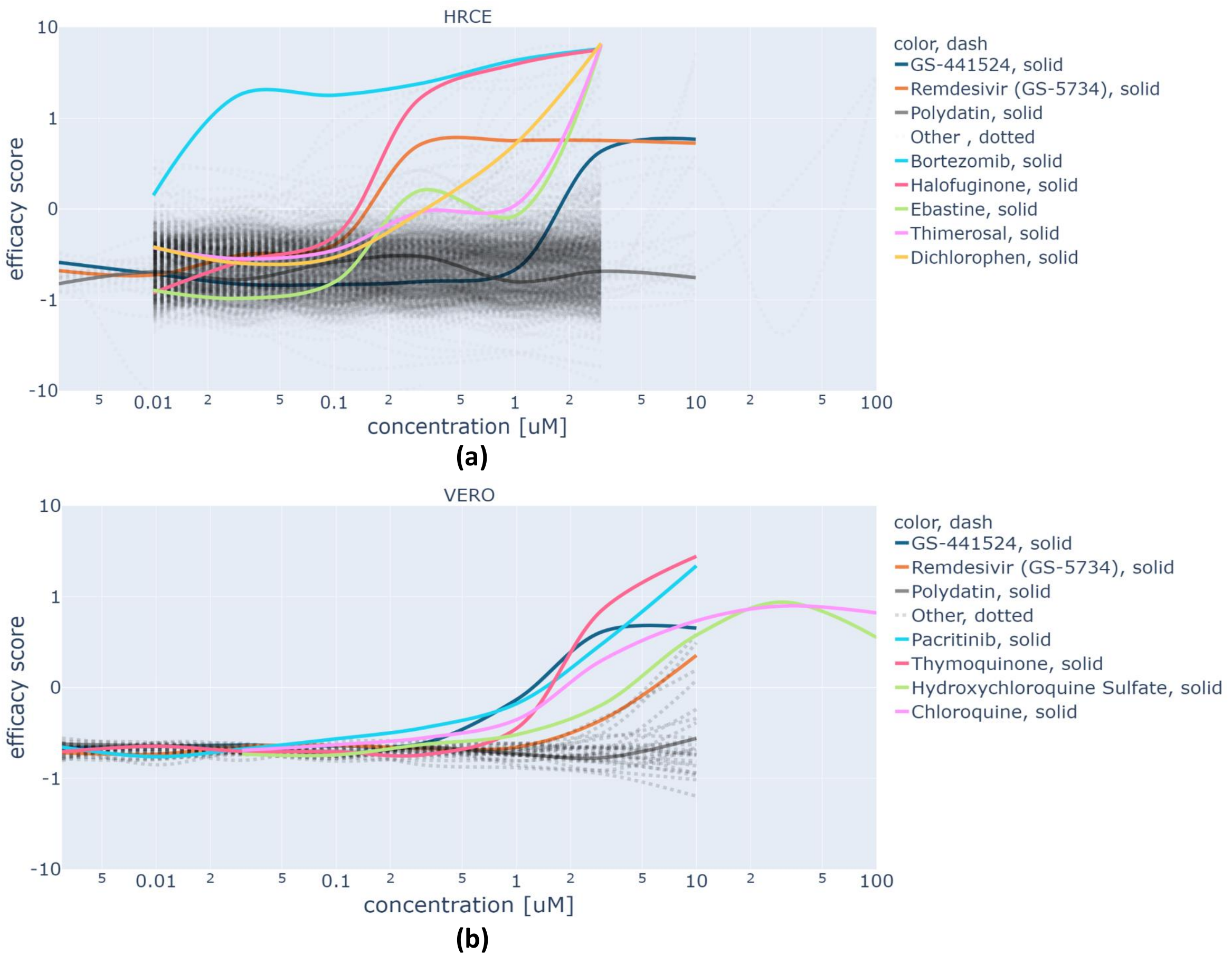}
\caption{
Dose response curves for HRCE (a) and VERO (b) cell lines obtained leveraging our GAN-DL's embedding.}
\label{fig:dose_response}
\end{figure}

As the figures of merit we propose: (i)~the embedding distributions, in the form of a scatter plot at varying concentrations, of \emph{Remdesivir} and \emph{GS-441524}, two compounds proved to be effective on SARS-CoV-2 in vitro in both the cell lines, and of \emph{Polydatin}, an inert compound that is known to be ineffective~\cite{baseline,ko2020comparative} (see \figureautorefname~\ref{fig:dose_response_scatter}). These compounds are shown as representative examples for both our embedding (a) and the baseline embedding (b); (ii)~the dose-response curves of a number of other compounds, obtained by reporting the corresponding mean \emph{efficacy score} at each concentration (see \figureautorefname~\ref{fig:dose_response}).

From \figureautorefname~\ref{fig:dose_response_scatter}, we can draw a number of considerations.
For the effective compounds \emph{Remdesivir} and \emph{GS-441524}, it is possible to see that progressively higher drug concentrations (corresponding to progressively darker red points in the scatter plots) tend to cluster towards positive values of the \emph{On-perturbation} axis in the $\textbf{E}^2$ space, closer to the region associated to the $C^+$group: the higher the concentration, the higher the \emph{On-perturbation} value. This is generally true for both the GAN-DL and the baseline embedding (see sections (a) and (b) of the figure, respectively), meaning that GAN-DL is equally able to represent the concentration-dependent ability of an active compound to preserve cell viability and inhibit SARS-CoV-2 infection. 

Differently from the effective compounds, the inert ones should reasonably behave the same in terms of SARS-CoV-2 inactivation, independently of their concentration. When looking at the plot of \emph{Polydatin} (a compound with no known effect on the virus in vitro), the values cluster towards the left side of the on perturbation axis where $C^-$ controls are located and do not show any specific color-pattern at increasing values of dose concentration. This demonstrates that, same as for the baseline, with GAN-DL embedding the ineffective compounds do not show any specific dose-dependent behaviour. Accordingly, very few values of the inert compounds are located in the positive \emph{On-perturbation} space (slightly greater then zero), suggesting no inactivation effect for SARS-CoV-2 infection in vitro.


While \figureautorefname~\ref{fig:dose_response_scatter} shows representative examples of compounds whose effectiveness in both cell lines is a-priori known~\cite{baseline,ko2020comparative}, \figureautorefname~\ref{fig:dose_response} reports the dose-response curves obtained with GAN-DL for all the screening compounds, of either known or unknown effectiveness. For both HRCE (a) and VERO (b), this figure shows on the x-axis the concentration values and on the y-axis the \emph{efficacy score} of the different compounds, as previously defined at the beginning of this section. To obtain the efficacy score axis, we normalize the \emph{On-perturbation} values using the controls, so that the mean of the negative controls is -1 and the mean of the positive controls is 1. By doing so, we obtain that the 0 value represents the \emph{efficacy threshold}, i.e. the value above which a compound is considered effective against SARS-CoV-2 infection in vitro. This normalization is performed on each cell line independently, as in the work by Cuccarese et al.~\cite{baseline}.

The curves of the three representative compounds shown in \figureautorefname~\ref{fig:dose_response_scatter} are also shown in \figureautorefname~\ref{fig:dose_response}, with solid colored lines to highlight them: \emph{GS-441524} (blue, solid), \emph{Remdesivir} (orange, solid) and \emph{Polydatin} (grey, solid). As it can be gathered from the figure, from a certain concentration value the curves of \emph{GS-441524} and \emph{Remdesivir} are above the efficacy threshold of zero. As the two cellular model behave differently upon SARS-CoV-2 infection, the concentration level above which a compound is effective is specific for the considered cell line.
This is an expected typical trend for an effective compound. On the contrary, the \emph{efficacy score} curves of \emph{Polydatin} are always below the value of zero, regardless the tested concentration. This confirms the expected ineffectiveness of the compound.
Besides \emph{GS-441524}, \emph{Remdesivir} and \emph{Polydatin}, \figureautorefname~\ref{fig:dose_response} shows solid colored lines also for the five compounds that obtained the highest efficacy scores in our screening. \emph{Bortezomib, Halofuginone, Ebastine, Thimerosal, Dichlorophen} tested the most effective in HRCE, while \emph{Pacritinib, Thymoquinone, Hydroxychloroquine Sulfate, Chloroquine} in VERO cells. 
For the sake of readability, all the remaining curves, associated with all the other tested compounds, are reported dashed grey and without a corresponding label.


In general, we can identify three different behaviors: i) under-threshold curves showing no specific correlation between concentration and efficacy score, same as \emph{Polydatin}; ii) almost-monotonically increasing dose response curves, featuring a positive correlation between concentration and efficacy: this is the most expected behavior for an effective compound, where a treatment requires a minimum dose to be efficacious; iii) dose-response curves that are above the efficacy threshold, but start decreasing after achieving a maximum efficacy score at a certain concentration value (see for instance \emph{Hydroxychloroquine Sulfate}, green solid line for the VERO cells). This is the case of a few compounds that were tested at high concentration values (100 uM). Hence, the drop of efficacy score can be reasonably explained by a loss of viability of the cell line related to a toxic effect of the compound at that high concentration.

\subsection*{Zero-shot learning}
\begin{figure}[h!]

    \begin{subfigure}[hb]{0.45\textwidth}
        \centering\includegraphics[width=0.89\textwidth]{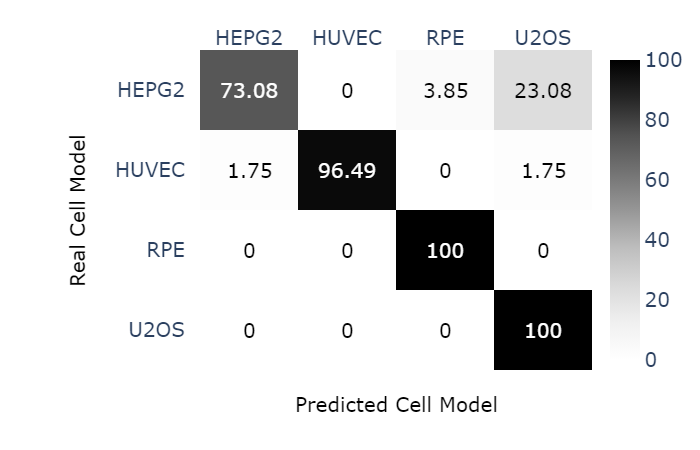}
        \label{fig:confusion_matrix_a}
        \caption{GAN-DL}
    \end{subfigure}
    \hfill
    \begin{subfigure}[hb]{0.45\textwidth}
        \centering\includegraphics[width=0.89\textwidth]{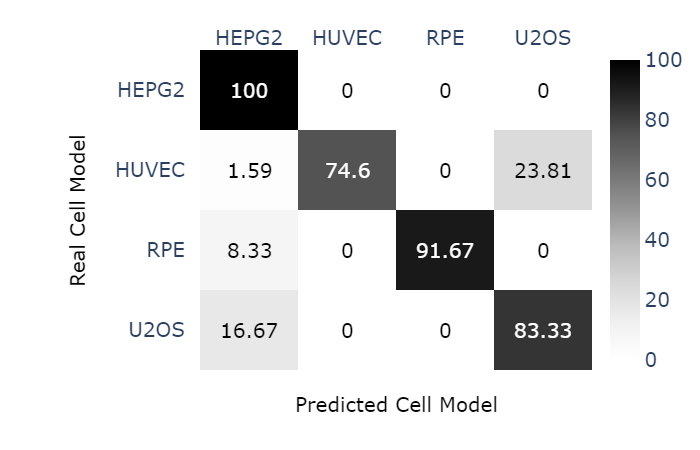}
        \label{fig:confusion_matrix_b}
        \caption{ImageNet pre-trained CNN}
    \end{subfigure}
    \caption{Confusion matrix of the zero shot cell classification task on the RxRx1 dataset.}
    \label{fig:confusion_matrix}
\end{figure}

In the previous subsections, we demonstrated that the proposed GAN-DL is able to characterize the two distinct cell lines included in RxRx19a dataset and to encode the dose-dependent information, even though it was not specifically trained for those tasks. Here, we try to assess the generalization capabilities of the model in a zero-shot learning experiment, that consists in a classification problem where at test time a learner observes samples from classes (i.e. cell lines) that were not observed during training. 
For this purpose, we exploit the RxRx1 image collection, a non-SARS-CoV2 related dataset consisting in 125,510 fluorescent microscopy images featuring human liver cancer cells (HEPG2), human umbilical vein endothelial cells (HUVEC), retinal pigment epithelium cells (RPE) and human bone osteosarcoma epithelial cells (U2OS) exposed to different perturbations~\cite{RxRx1}(i.e. 1,108 different siRNAs to knockdown 1,108 different genes). For the sake of channels compatibility, to perform a zero-shot inference on the RxRx1 dataset we removed the channel corresponding to the MitoTracker, a dye that stains mitochondria, that is not present in the five-staining protocol of RxRx19a.
Same as in the previous experiments, we exploit a soft margin linear SVM built on top of our GAN-DL embedding to categorize the four different cell lines included in the RxRx1 benchmark. We show the corresponding results in the form of a confusion matrix in \figureautorefname~\ref{fig:confusion_matrix}(a). From this matrix we can see that, despite the fact that the RxRx1 cell lines are totally new for GAN-DL (i.e. the backbone StyleGAN2 was not trained to generate the RxRx1 cellular images but the ones of RxRx19a), they can be linearly separated in the feature space with a mean accuracy of 92.68\%. This is not much lower than the accuracy that was obtained on the RxRx19a dataset (see \tableautorefname~\ref{tab:dowstream_task}).

For comparison, we show the  results obtained by a DenseNet classifier (the same architecture of our previous experiment), pre-trained on ImageNet and fine-tuned on the RxRx1 dataset. In spite of the fine-tuning, as shown in the confusion matrix of \figureautorefname~\ref{fig:confusion_matrix}(b), the DenseNet classifier obtained an accuracy of 83.19\%, about 10\% lower than GAN-DL.

\section*{Methods}
\label{sec:method}
\subsection*{Dataset}
The data used in this work are part of the RxRx datasets collections, that are available online~\cite{rxrx}. More specifically, in our experiments we exploit:
\begin{enumerate}
    \item The RxRx19a, which, as briefly mentioned in Results \sectionautorefname,~gathers several experiments aimed at investigating therapeutic potential treatments for COVID-19 from a library of FDA-approved and EMA-approved drugs or compounds in late-stage clinical trials\cite{RxRx19}. After 24 hours post-seeding, the cells have been infected with SARS-CoV-2 and then incubated for 96 hours before fixation, staining and imaging. Images were produced using five channels to highlight the cell membrane and different cellular compartments, leveraging a specific fluorescent staining protocol, as described in the work by Cuccarese and colleagues\cite{RxRx19}. The compounds were screened by treating cells in six half-log doses with six replicates per dose for each compound approximately two hours after cell seeding. Further details about the assays protocol can be found at the official dataset website \cite{rxrx}. The resulting dataset is made up of 305,520 fluorescent microscopy images of size equal to $1024\times1024\times5$. To assess the specificity of the tested compounds, two suitable control groups have been designed. The first one consists in conditioned media preparations generated from uninfected cells (Mock), the second one is made up of cells infected in vitro by active SARS-CoV-2 virus and not treated with any compounds.
    \item The RxRx1, a dataset consisting of 296 GB of 16-bit fluorescent microscopy images, created under controlled conditions to provide the appropriate data for discerning biological variation in the common context of changing experimental conditions. The RxRx1 has been specifically created to push innovative machine learning and deep learning pipeline on large biological datasets, aimed at drug discovery and development\cite{RxRx1}.
\end{enumerate}

\subsection*{GAN-DL's backbone: the StyleGAN2 model}
The recent literature about GANs is focused on methodologies to improve their training and counteract the well known difficulties and limitations of this phase~\cite{R1}. More specifically, Wasserstein Generative Adversarial Networks (W-GANs)~\cite{arjovsky2017wasserstein} have been introduced to prevent two common problems of training GANs. First, mode collapse is a form of GAN failure in which the network learns to generate only a subset of the data, eventually a single image. The discriminator get trapped into a local minimum and the generator easily presents the same examples over and over to convince the discriminator. This results in a model that is heavily over-fitted on this particular subset. Second, lack of convergence due to either the generator or the discriminator, which are improving at a faster pace than the other network. This prevents the mutual improvement that is necessary for convergence. 

\begin{figure}[!htbp]
    \captionsetup[subfigure]{justification=centering}
    \centering
    \begin{subfigure}[t]{0.12\textwidth}
        \includegraphics[width=\textwidth]{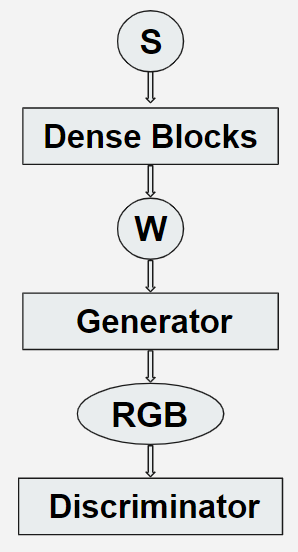}
        \caption{High-level overview}
    \end{subfigure}
    \hfill
    \begin{subfigure}[t]{0.30\textwidth}
        \includegraphics[width=\textwidth]{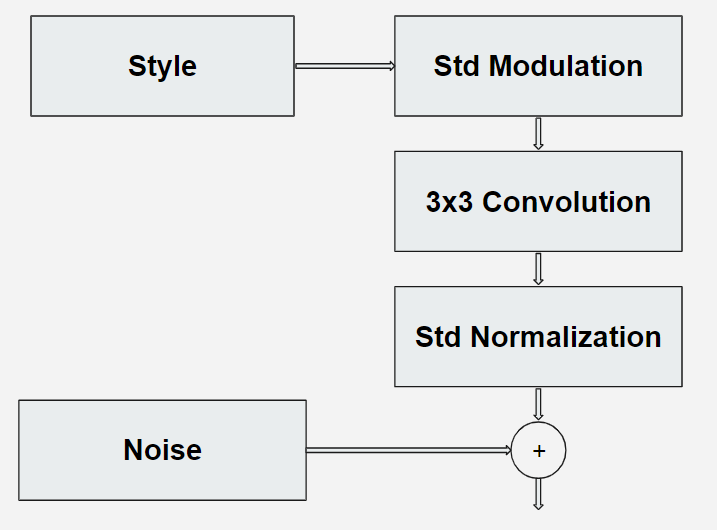}
        \caption{Generator Block}
    \end{subfigure}
    \centering
    \begin{subfigure}[t]{0.275\textwidth}
        \includegraphics[width=\textwidth]{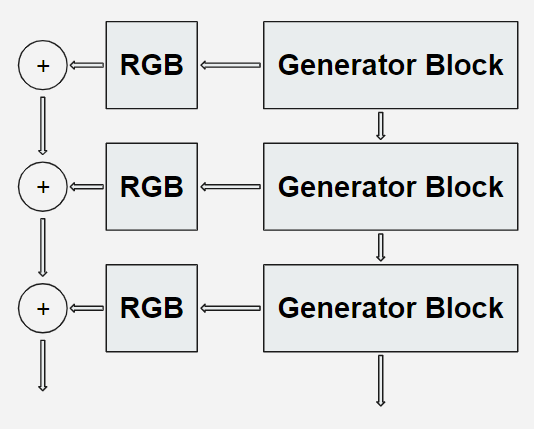}
        \caption{Residual Connections (Generator)}
    \end{subfigure}
    \begin{subfigure}[t]{0.24\textwidth}
        \includegraphics[width=\textwidth]{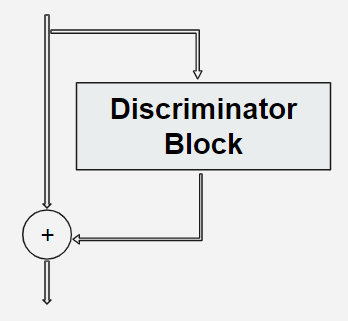}
        \caption{Residual Connections (Discriminator)}
    \end{subfigure}
    
    \caption{Overview of the StyleGAN2 architecture}
    \label{fig:StyleGAN2}
\end{figure}

W-GANs  have proved to be an efficient solution to overcome both those limitation at once, by replacing the classical discriminator model with a critic that scores the realness of a given image by means of the so-called Wasserstein distance~\cite{arjovsky2017wasserstein}. For our GAN-DL we employed the Nvidia's StyleGAN2 architecture~\cite{styleGAN2}, that is an instance of W-GAN with recurrent connections in both the generator and the discriminator. \figureautorefname~\ref{fig:StyleGAN2} shows a high level diagram of this architecture (a), as well as a breakdown of the generator block and residual connections (b-c). We refer the reader to the paper of Karras et al~\cite{styleGAN2} for technical details.

The original StyleGAN2 model has been scaled down to allow training on more reasonable hardware and time-frames. To reduce the number of parameters, we simplified the fully connected mapping network to be 3 layers deep instead of the original 8. The latent space we employ corresponds to the style vector, the sizing of which is 512 in accordance with the original paper, while the latent space of the other embeddings shown for comparison is of size 1024 (more details in Counterpart embeddings \subsectionautorefname).

\subsection*{Experimental setup}
The StyleGAN2 backbone was trained on the RxRx19a dataset using Adam optimizer with a learning rate of \(10^{-4}\), with the same loss as the one described in the StyleGAN2 paper\cite{styleGAN2}. No hyperparameter optimization was performed. Conversely, we employed two regularization terms:
\begin{itemize}
    \item Generator: Jacobian Regularization (also known as PPL regularization) \cite{styleGAN2}, Exponential Moving Average of the weights \cite{proGAN}
    \item Discriminator: Lipschitz L1 penalty \cite{L1}, R1 regularization \cite{R1}
\end{itemize}

For training we employed one TPU v3-8 node with 16GiB of RAM per core. TPUs are AI accelerator ASICs (Application Specific Integrated Circuits) which have the ability to train neural networks significantly faster than GPUs by executing a larger amount of computations in parallel.

The original StyleGAN2 took 9 days on 8 Tesla V100 GPUs to train on the FFHQ dataset, while our slimmed and repurposed version required 24 hours on a TPU v3-8 node or 48 hours on a single Tesla V100 GPU to obtain the results shown in this paper. Most of the difference in training time can be attributed to the vastly different dataset used as well as the different training objective.

\subsection*{Counterpart embeddings}
\label{sec:otherembeddings}
In our experiments, GAN-DL embedding is compared against three different counterparts:
\begin{itemize}
    \item The RxRx19a embedding, released together with the imaging data by Cuccarese et al~\cite{baseline} and referred to as \emph{baseline} in this manuscript (see \figureautorefname~\ref{fig:clustering}, \figureautorefname~\ref{fig:dose_response_scatter} and \tableautorefname~\ref{tab:dowstream_task}). It consists of 1024-dimensional vectors (one vector per image) obtained using a DenseNet CNN architecture as the backbone, pre-trained on a source labelled dataset with similar imaging characteristics (RxRx1 dataset). The proprietary model is not publicly released by the authors.
    \item The embedding of a DenseNet CNN pre-trained on a source dataset with completely different imaging characteristics and contents (ImageNet) and fine-tuned on a labelled portion of the target RxRx19a dataset, i.e. the controls (see~\tableautorefname~\ref{tab:dowstream_task}). For a fair comparison, the backbone of this methodology is a DenseNet, same as for the baseline solution. 
    \item The embedding of a DenseNet CNN pre-trained on a source dataset with completely different imaging characteristics and contents (ImageNet) and fine-tuned on the RxRx1 dataset (see~\figureautorefname~\ref{fig:confusion_matrix}). 
\end{itemize}

Note that pre-training a neural network with ImageNet data involves interpreting images in terms of RGB channels, while fluorescent cellular images such as the RxRx19a and RxRx1 datasets are usually represented in 5/6 channels. To account for this difference, we introduce a trainable convolutional layer with a kernel size of 1 at the beginning of the RGB pre-trained networks, so that the fluorescent images are converted to 3 channels.

\section*{Conclusions}
\label{sec:discussion}

In contexts where dataset annotation is costly, like medical and computational biology domains, the current standard, for the application of deep learning models on image data, involves the use of a ImageNet-pretrained CNN model, and optionally fine-tuned on the limited quantity of labelled samples that are available for the given application. Nevertheless, we found such transfer learning-based strategy totally unsatisfactory for our real word application (see \tableautorefname~\ref{tab:dowstream_task}), where the inherent complexity of the required biological tasks and the experimental set-up of a large scale drug screening initiative claims for a more powerful representation learning technique. If, in general, SSRL seems a promising solution for those scenarios suffering a paucity of labelled data, the recent work by Wallace et al.\cite{wallace2020extending_SSRL} has shown how traditional SSRL featurization methodologies fail in several biological downstream tasks. This is mainly imputed on the difficulty in defining a pretext task which can be exploited by traditional contrastive SSRL.

On top of these considerations, in this paper we propose GAN-DL, a fully SSRL method leveraging the representation learning acquired by the discriminator of a StyleGAN2 model \cite{styleGAN2}. Our GAN-DL does not require any task-specific label to obtain the image embedding, as the StyleGAN2 backbone is trained on a generative task based on the competition of a generator and of a discriminator, that is completely independent on the downstream task. By doing so, we address the problem of lack of annotated data, that is instead necessary for conventional CNN-based transfer learning methods. We demonstrated the goodness of our featurization methodology in two downstream supervised tasks: the classification of different cellular models (HRCE versus VERO cells) and the categorization of positive versus negative control groups in the RxRx19a benchmark \cite{RxRx19}. For this purpose, we trained a simple linear SVM on top of the self-supervised GAN-DL embedding, which does not require a large number of annotated data. Furthermore, we compared our solution with a baseline state-of-the-art DenseNet121 model, pre-trained on the RxRx1 dataset\cite{RxRx1} (the corresponding embedding is released together with the imaging data by Cuccarese et al. \cite{baseline}). 

On the one hand, the baseline embedding is generally more accurate than GAN-DL in the downstream classification tasks, even though by a small margin. On the other hand, the baseline is pre-trained on a very large annotated dataset (RxRx1 dataset, consisting of 296 GB of fluorescent microscopy images), while training GAN-DL does not require any task-specific image annotations. This is indeed a major advantage for the re-usability of our method in different contexts where annotated data from a similar domain are few or even not available at all, which is a frequent challenge of many biological applications.

We speculate that our GAN-DL embedding, leveraging as pre-text task the generation of plausible and high resolution images through the adversarial game between the generator and the discriminator, proficiently learns an unbiased and disentangled image featurization able to describe the fine-grained patterns that are typical of biological applications. This leads to an improved capability of separating multiple unrelated features along different axis of the latent space, which should be ultimately helpful to address any downstream tasks requiring knowledge of the salient attributes of the data \cite{chen2016infogan}. To demonstrate our claim, we put this capability of GAN-DL into effect in a number of different applications:
(i) the classification of active and inactive compounds against SARS-CoV-2 infection in two different cell lines; (ii) the generation of dose-response curves for the large scale molecule screening of RxRx19a, without the need of any training on purposely labelled data; (iii) the zero-shot learning of four different cell lines included in the RxRx1 dataset. The satisfactory results in all the presented scenarios demonstrate the goodness and generalization capability of our approach and legitimize the future exploitation of generative SSRL even in other biological applications.

\bibliographystyle{unsrt}
\bibliography{biblio.bib}  
\end{document}